# Blockchain based Spectrum Sharing

Chen SUN [†], Shuo WANG [†]

† Wireless Network Research Department
Beijing Laboratory, R&D Center China, Sony (China) Limited
Room 301, 3/F, Commercial Podium, Ping An international Financial Center
No.1 South Xinyuan Rd., Chaoyang District, Beijing. 100027 P. R. C
E-mail: chen.sun@sony.com

**Abstract** Spectrum sharing has long been considered as method to improve spectrum resource utilization. Centralized geolocation database approach has been accepted globally for commercial applications. Recently blockchain has been considered as a platform to support spectrum sharing in a distributed manner. Like other commodities, spectrum or right of spectrum usage can be exchanged through blockchain. However, this leads to changes of the location where a particular frequency channel is utilized which brings potential interference. In this paper, we present the interference-based consensus and smart contract for blockchain based spectrum sharing. We will also introduce the other aspects of blockchain such as cross-chain transaction, cross-tier spectrum coordination and incentive mechanism that need to be studied for spectrum sharing applications.

**Keyword** Cognitive radio system (CRS), dynamic spectrum access (DSA), coexistence, blockchain, consensus, smart contract

## 1. Introduction

With the increasing bandwidth demands of future wireless services, spectrum regulators face great challenge in allocating finite spectrum resources. However, the licensed spectrum is often underutilized temporally and spatially. To tackle this problem, a novel paradigm called spectrum sharing is proposed, which give the opportunity for secondary systems to use unoccupied spectrum resources owned by a primary system. The existing spectrum management technologies such as TVWS and CBRS use centralized GLDB to control spectrum assignment for wireless devices and protect incumbent users [1][2]. Although the centralized solutions are effective for dynamic spectrum access, they require high computation capability for spectrum managers because all the interference calculation and spectrum assignment decisions are made by a central entity. Recently, blockchain based decentralized approaches are proposed for spectrum management especially in China and Europe.

Blockchain is a type of distributed ledger technology that organizes data into blocks, which are chained together in an append-only mode. A consensus mechanism is executed to validate transactions among peer nodes and keep the ledger on each node updated. It has the characteristics of decentralization, anonymity, and auditability.

Some characteristics of spectrum sharing are consistent with the application scenario of blockchain. Firstly, spectrum usage is real-time and requires timely processing, which can be enabled by localized spectrum ledger. Secondly, spectrum sharing involves multiple stakeholders that may not trust each other, which can be solved by blockchain. Thirdly, with blockchain, spectrum sharing can be conducted among wireless devices directly without the need for intermediaries.

However, there are uniqueness of spectrum trading comparing to trading of other commodities. The most important aspect is that transaction of spectrum means changes of locations of the radio transmitter. This brings potential interference to those "innocent" wireless networks nearby the sellers and buyers. In a tiered system, spectrum usage is bounded by the interference budget at the primary system. Trading might involve the consumption of interference budget other than direct exchanging frequency bands. Another aspect is that the variations of wireless channels used for block propagation may affect the performance of blockchain systems compared with applications in wired environment.

This paper introduces various aspects of blockchain based spectrum sharing such as the interference-based consensus, block design, cross-chain transaction, cross-tier spectrum coordination and incentive mechanism. The paper is organized as follows. Section 2 presents the standardization activities of the spectrum sharing and blockchain technology. The technical aspects of blockchain based spectrum sharing are given in Sections 3. Finally, Section 4 concludes the paper.

## 2. Standardization Activities of Spectrum Sharing and Blockchain

### 2.1. Standardization of Spectrum Sharing

The Television white space (TVWS) technology is firstly standardized for spectrum sharing. Several IEEE working groups such as IEEE 802.11af, 802.22, 802.19.1 and P1900.6 have developed standards on different aspects of TVWS including system requirements, architecture, sensor information exchange and so forth. [3] [4].

In the United States, the wireless innovation forum (WinnForum) has developed standards for spectrum sharing in CBRS band which is 3550-3700 MHz. In Europe, ETSI has developed the standards of mobile broadband services under LSA in the 2300-2400 MHz band.

### 2.2. Standardization of Blockchain Technology in Communications

Multiple study groups in the international telecommunication union (ITU), such as FG DLT，SG13 and SG20, have been working on the standardization of blockchain technology including the concepts, use cases and reference architecture of DLT-based applications and services. ETSI established an industry specification group (ISG) on permissioned distributed ledger (PDL) in 2018. The IEEE standard association has been making efforts on blockchain standardization with various activities such as DLT use in Internet of things (IoT) data management, connected automated vehicles, blockchain access control and blockchain interoperability.

The China Communications Standards Association (CCSA) has been conducting blockchain related study items and work items in multiple working groups. The technical report on blockchain and applications in next generation wireless network identified spectrum sharing as one of the important use cases of blockchain technology [5]. Then, the research on blockchain based solutions for wireless network architecture further discussed the key issues and technical requirements of blockchain based dynamic spectrum sharing [6].

## 3. Key Issues and Potential solutions for Blockchain based Spectrum Sharing

### 3.1. Consensus mechanism considering interference

One of the most important differences between spectrum sharing and other blockchain based data sharing is the impact of interference. In wireless networks, the validation speed needs to be improved to ensure quality of services (QoS). Our proposed approach in [7] is to select a few nodes in a validation area based on the interference relationship. The transaction is considered valid and will be executed only if all the nodes in the validation area approve it showing that they accept the interference that will be incurred due to the transaction. The small validation zone also allows to save consensus time comparing to traditional scheme, such as Bitcoin, which populates the transaction data to all nodes in the network.

### 3.2. Incentive Mechanism

In a scenario where multiple mobile network operators (MNO) and service providers (SP) utilize a blockchain based spectrum trading platform to manage spectrum usage and trading information. MNOs (as leaders) can sell spectrum to SPs to achieve economic benefits while SPs (as followers) can provide better services for end users. How to reach a balance among these stakeholders? Due to the competition relation among SPs and the selfishness of maximizing their own benefits, we model the spectrum trading process as a hierarchical Stackelberg game and propose a compensation-incentive mechanism to maximize the utility of spectrum buyer and ensure fairness among competitive SPs [8].

### 3.3. Cross-chain Transaction

The throughput of single-chain structure limits the performance of large-scale spectrum trading. We propose a multi-blockchain architecture and a corresponding cross-chain mechanism to improve the speed of spectrum trading [9]. As is mentioned before, spectrum transactions cause the change of interference relationships among incumbents and wireless devices. Therefore, it is necessary to find a trusted and reliable intermedia to ensure the security of cross-chain spectrum trading. we further combine the notary and relay-chain mechanism in this architecture, where coexistence managers play the roles of the notary and the selected CBSDs will maintain a decision blockchain. The decision blockchain is used to share information between different groups and can improve the scalability of notary mechanism.

### 3.4. Cross-tier Spectrum Coordination

Some spectrum sharing frameworks such as CBRS have multiple tiers of different priorities. To protect incumbent users, the aggregated interference from secondary users of all tiers should not exceed an interference threshold. Currently, this interference budget is allocated equally or in a fixed proportion among all secondary users. Blockchain technology provides a promising solution for coordination of spectrum usage among different tiers. Firstly, a spectrum blockchain is established for the devices sharing the same spectrum. The devices in the lower priority tier can access to the spectrum only if they will not cause harmful interference to upper tiers and incumbent users. Through the spectrum blockchain, devices in lower tier can request the devices in upper layer to adjust their interference budget and ensure the aggregated interference to incumbent users does not exceed the protection threshold. This cross-tier spectrum coordination can improve the total number of lower tier users which can have access to the shared spectrum.

## 4. Conclusion

In this article, the standardization efforts of spectrum sharing and blockchain technology by various standards developing organizations were summarized. Then, the key issues in blockchain based dynamic spectrum sharing were presented and solutions to these issues were proposed. Through the discussion, we see great potential for the application of blockchain technology in dynamic spectrum sharing, which will pave the way for efficient and fair utilization of valuable spectrum resources for various wireless network such as mobile public networks, Wi-Fi, V2X networks and private networks.